\documentstyle[12pt,twoside]{article}
{\catcode `\@=11 \global\let\AddToReset=\@addtoreset}
\AddToReset{equation}{section}

\def\greaterthansquiggle{\raise.3ex\hbox{$>$\kern-
.75em\lower1ex\hbox{$\sim$}}}
\def\lessthansquiggle{\raise.3ex\hbox{$<$\kern-
.75em\lower1ex\hbox{$\sim$}}}
\newcommand{\beq}{\begin{equation}}
\newcommand{\eeq}{\end{equation}}
\newcommand{\beqa}{\begin{eqnarray}}
\newcommand{\eeqa}{\end{eqnarray}}
\newcommand{\beqan}{\begin{eqnarray*}}
\newcommand{\eeqan}{\end{eqnarray*}}
\newcommand{\ba}{\begin{array}}
\newcommand{\ea}{\end{array}}
\newcommand{\no}{\nonumber}

\newcommand{\A}{{\cal A}}

\newcommand{\C}{{\cal C}}

\newcommand{\F}{{\cal F}}

\newcommand{\Z}{{\cal Z}}

\def\nz{\ifmmode {I\hskip -3pt N} \else {\hbox {$I\hskip -3pt N$}}\fi}
\def\zz{\ifmmode {Z\hskip -4.8pt Z} \else
       {\hbox {$Z\hskip -4.8pt Z$}}\fi}
\def\qz{\ifmmode {Q\hskip -5.0pt\vrule height6.0pt depth 0pt
       \hskip 6pt} \else {\hbox
       {$Q\hskip -5.0pt\vrule height6.0pt depth 0pt\hskip 6pt$}}\fi}
\def\rz{\ifmmode {I\hskip -3pt R} \else {\hbox {$I\hskip -3pt R$}}\fi}
\def\cz{\ifmmode {C\hskip -4.8pt\vrule height5.8pt\hskip 6.3pt} \else
       {\hbox {$C\hskip -4.8pt\vrule height5.8pt\hskip 6.3pt$}}\fi}
\def\au{{\setbox0=\hbox{\lower1.36775ex%
\hbox{''}\kern-.05em}\dp0=.36775ex\hskip0pt\box0}}
\def\ao{{}\kern-.10em\hbox{``}}

\renewcommand{\baselinestretch}{1.5} 
\normalsize
\begin{document}
\bibliographystyle{plain}

\begin{titlepage}
\begin{flushright}
UWThPh-2001-33 \\

\end{flushright}
\vspace{2cm}
\begin{center}
{\Large \bf  Quantizing Yang--Mills Theory on a 2-Point Space }\\[40pt]
$ \mbox{Helmuth H\"uffel}$\\
Institut f\"ur Theoretische Physik \\
Universit\"at Wien \\
Boltzmanngasse 5, A-1090 Vienna, Austria
\vfill

{\bf Abstract}
\end{center}
\renewcommand{\baselinestretch}{1.0} 
\small

We perform the Batalin-Vilkovisky quantization of Yang--Mills theory on
a 2-point 
space, discussing  the formulation of Connes--Lott as well as  
Connes' real spectral triple approach. Despite of the model's apparent
simplicity 
the gauge structure
reveals infinite reducibility 
and the gauge fixing is afflicted with the Gribov problem.

\vfill

\end{titlepage}
\renewcommand{\baselinestretch}{1.5} 
\normalsize

\section{Introduction}

Noncommutative geometry constitutes one of the  fascinating new concepts 
in current
theoretical physics research with many promising impacts and
applications in a
diverse set of fields \cite{Connes,Mad,Landi,GracVar,Kastler}.
Specifically we
 mention the construction of the classical action of the standard 
model
 \cite{gravity,spectral}, unifying the Einstein--Hilbert action, the
 Yang--Mills action, the  
 Dirac action, and  the Klein--Gordon action with the 
 Higgs potential and 
 spontaneous symmetry breaking.  
 
 The basic idea of noncommutative 
 geometry is to replace the notion of  differential manifolds  and
functions 
 by specific    
 noncommutative  
 algebras of
 functions; 
 the  geometric setting of  gauge theories  
 as fibre bundles finds a 
 noncommutative 
 generalization
 in terms of finitely 
 generated projective modules over  noncommutative algebras.

 It seems, however, that within this noncommutative algebraic
 framework  the 
 concepts of {\it quantizing} gauge theories, in specific the issue of 
 gauge fixing and the proper definition of  
  a  path integral measure for the standard model
 are not yet fully understood \cite{Trieste}. 
 Our intention  for this paper is not to present new results 
 in these rather fundamental issues.  Instead   we quantize  
 one of the simplest toy 
 models for noncommutative gauge theories, which is Yang--Mills theory 
 on a 2-point space, by applying the standard 
 Batalin--Vilkovisky   method \cite{BV,BV2,Henn,Gomis}. 
Somewhat surprisingly we find that despite of the model's  original
simplicity 
the gauge structure 
reveals infinite reducibility 
and the gauge fixing is afflicted with the Gribov \cite{Gribov} problem.

 In section 2 we  work out the formulation of the model following the 
approach of Connes--Lott \cite{ConnL}. In 
section 3 the infinite reducibility of the gauge symmetry is explained;
the 
Batalin--Vilkovisky quantization  of the model is 
performed in section 4. We discuss the Gribov problem  in section 5
and finally, in section 6,  recast our results within Connes' 
real spectral triple 
approach \cite{gravity,reality}.

\section{The Formulation of
Connes--Lott}

Following \cite{ConnL} we define the Yang--Mills Theory on a 2-point space
in terms of the algebra ${\mathbf A}=C \oplus C$, which is represented 
 by diagonal  complex valued $2\times 2$ 
 matrices; the
Dirac operator $D$ is given by \mbox{$D=\left( \ba{cc}
0 & \mu  \\
 \mu   & 0 
\ea \right)$}, where $\mu \in \rm R$ is an arbitrary  parameter. 
The differential p-forms $\omega_{p}$ are constant, diagonal or
offdiagonal  $2\times 2$  matrices, 
depending on whether p is even or odd,
respectively. One  has a $\Z_{2}$ grading of matrices (to be diagonal or 
 offdiagonal) and obtains
  a matrix derivative $\bf d$. Acting on   
$2\times 2$ matrices it
is a nilpotent graded derivation\footnote
{ \qquad ${\bf d}\,a =i \mu\left( \ba{cc}
a_{21}+a_{12} &  a_{22}-a_{11}  \\
a_{11}-a_{22} & a_{21} +a_{12}
\ea \right) 
\qquad {\rm where } \qquad
a = \left( \ba{cc}
a_{11} & a_{12}  \\
a_{21} & a_{22} 
\ea \right), \quad a_{ij} \in \rm C. 
$
} with respect to the matrix product 
and the matrix $\Z_{2}$ grading.

Specifically   the  
1-forms  are given by $\omega_{1}=a \,{\bf d} b$, where $a,b 
\in \,\mathbf{A}$,  which  are odd (i.e. offdiagonal) matrices. The
subset of 
anti-Hermitean 1-forms $\A$  can be 
parametrized by
\beq
\A = \left( \ba{cc}
0 & i \mu \phi  \\
i \mu \bar \phi  & 0 
\ea \right) 
\eeq
and constitute the gauge 
fields of the model; here $\phi\in \rm C$  
denotes a 
(constant)
scalar field.
The (rigid) gauge 
transformations of $\A$ are defined by
\beq
\label{old}
\A^{U}=U^{-1}\A U+U^{-1}{\bf d}\,U
\eeq
with $U$ being  a unitary element of the algebra $\mathbf{A}$. It is a 
constant, 
 even  and unitary matrix which we define to 
have only 
abelian entries; it can exponentially be 
parametrized by the even   matrix 
$\varepsilon$
\beq
U = \left( \ba{cc}
 e^{i \alpha} & 0  \\
0 & e^{i \beta} 
\ea \right)=e^{i \varepsilon}, 
\quad \varepsilon=\left( \ba{cc}
\alpha & 0  \\
0 & \beta 
\ea \right)\quad\alpha,\beta \in \rm R.
\eeq
We point out that the Yang--Mills theory on the
2-point space is an ideal play ground to study quantization 
techniques: Due to the nonabelian form of the gauge transformations 
(\ref{old}) the model shares many interesting features with the standard
 Yang--Mills 
theory, yet   it has no physical 
space-time dependence and allows  extremely simple  calculations. We 
even  restrict 
ourselves to just abelian entries along the diagonal of $U$, thus   
studying a  $U(1)\times U(1)$ gauge model with nonabelian features.

We define a 
scalar product for $2\times 2$ matrices $a,b$ by $\langle 
a \,\vert\, b \rangle=tr\, a^{\dagger} \,b$ where $\dagger$ denotes
taking the  Hermitian conjugate. 
The curvature $\F$ is defined as usual by $\F={\bf d}\A+\A\,\A$ and 
transforms under gauge transformations as $\F^{U}=U^{-1} \F U$; 
for an action which is  automatically invariant  under the gauge 
transformations (\ref{old})  one takes
\beq
\label{action}
S_{inv}=\frac{1}{2}\langle \F \vert \F \rangle.
\eeq
Written out in components  the  scalars' contribution is 
given by 
\beq
\label{components}
S_{inv}=\mu^4\left( (\phi+\bar\phi)+\phi \,\bar \phi\right)^2.
\eeq
It was pointed out in \cite{Sitarz93,Sitarz95} that the most general
form of the gauge 
invariant action also allows  a term proportional to $tr 
\F$
\beq
\label{totalaction}
{\hat S}_{inv}= \frac{1}{2} \Bigl( \langle  \F\,\vert\,\F\rangle
+\gamma\, 
tr \F \,\Bigr) 
\eeq
where $\gamma \in 
\rm R$ is an arbitrary parameter. 
We  note, 
however, that one requires the scalars $\phi$ to be vanishing at the 
minimum of 
the action so that in the case of (\ref{totalaction})  the scalars have
to be 
shifted  appropriately. 
Explicitly we have 
\beq
{\hat S}_{inv}=
\mu^4(2u+u^2+v^2)(2u+u^2+v^2-\frac{\gamma}{\mu^2})
\eeq
where we introduced $\phi= u + i v$. 
Whereas the local maximum is at
\beq
\label{max}
u_{max}=-1, \quad v_{max}=0
\eeq
the circle of local minima is given by 
\beq 
(u+1)^2+v^2=1+\frac{\gamma}{2 \mu^2},\quad {\rm where}\quad \gamma\geq 
-2 \mu^2.
\eeq
We   choose 
\beq
u_{min}=\sqrt{1+\frac{\gamma}{2 \mu^2}}-1, \quad v_{min}=0
\eeq
and define  shifted scalars $\tilde \phi=\phi-u_{min}$ which  by
construction  are 
vanishing at the minimum
of the action. 
From
\beq
{\hat S}_{inv}= \mu^4\left((\tilde \phi +\bar{\tilde 
\phi})\sqrt{1+\frac{\gamma}{2 \mu^2}} \, +\tilde \phi \bar{\tilde 
\phi}\right)^2-\frac{\gamma^2}{2}.
\eeq
we omit the irrelevant  constant $-\frac{\gamma^2}{2}$, rescale  $\tilde 
\phi=\hat\phi\sqrt{1+\frac{\gamma}{2 \mu^2}}$ and 
$\mu\sqrt{1+\frac{\gamma}{2 \mu^2}}=\hat\mu$ so that finally
\beq
{\hat S}_{inv}={\hat\mu}^4\left( (\hat\phi+\bar{\hat\phi})+
\hat\phi \,\bar{\hat\phi}\right)^2.
\eeq
We see that  the inclusion of the action term linear 
in $\F$ can  be compensated by shifting and rescaling of the scalar 
field
$\phi$, as well as  by rescaling of the  parameter 
$\mu$. 
As the scalar fields and the parameter  are  arbitrary from the outset 
the inclusion of the action term linear in $\F$ appears to be
unnecessary. In the following  we will set $\mu=1$ for simplicity and
stick to the 
action term (\ref{action}) quadratic in $\F$.

\section{Gauge Transformations and Infinite Reducibility}

The  (zero-stage) gauge transformations (\ref{old}) 
explicitly are given by
\beq
\A^{U}=\left( \ba{cc}
0 & i e^{i(\beta-\alpha)}(\phi+1)-i \\
ie^{-i(\beta-\alpha)} (\bar \phi+1) -i & 0
\ea \right), 
\eeq
so that  the usual abelian gauge transformations are 
implied for the Higgs fields $H=\phi+1$ and $\bar H=\bar\phi+1$.
To discuss  infinitesimal  (zero-stage) gauge transformations we
introduce 
an   even, infinitesimal (zero stage) gauge parameter matrix 
$\varepsilon^0_{e}$ in terms of which
 $U \simeq {\bf 
1}+\varepsilon^0_{e}$. The  infinitesimal (zero-stage) gauge variation
of $\A$
derives as 
\beq
\label{zero}
\delta_{\varepsilon^0_{e}}{\A}=i {\bf R}^0 \,\varepsilon^0_{e} \quad 
{\rm where} \quad
{\bf R}^0={\bf D};
\eeq
here the \mbox{(zero-stage)} gauge generator ${\bf R}^0$ is defined in
terms of the
 covariant matrix derivative ${\bf D}$, which acting on 
$\varepsilon^0_{e}$ is given by 
${\bf D}\varepsilon^0_{e} ={\bf 
d}\varepsilon^0_{e}+[\A,\varepsilon^0_{e}]$.

A gauge symmetry is called  irreducible if the  (zero 
stage) gauge generator ${\bf 
R}^0$ does not posess any zero mode  \cite{BV,BV2,Henn,Gomis}. 

It is amusing to note that the Yang--Mills theory on the 2-point 
space reveals an infinitely reducible gauge symmetry: We observe 
that
  ${\bf D\,d}$ is vanishing on arbitrary odd matrices. Thus
there 
exists a zero mode $\varepsilon^1_{e}$ for  the (zero-stage) gauge 
generator 
${\bf R}^0$, such that
\beq
{\bf R}^0\varepsilon^1_{e}=0 \quad{\rm where} \quad
\varepsilon^1_{e}={\bf R}^1
\varepsilon^1_{o} \quad {\rm with} \quad{\bf R}^1={\bf d}.
\eeq
Here $\varepsilon^1_{o}$ denotes    an odd, infinitesimal (first-stage) 
gauge parameter matrix and
  ${\bf R}^1$ the corresponding (first-stage) gauge generator.
As a matter of fact  
an infinite tower of  (higher-stage) gauge generators
${\bf R}^s$,\,\,\mbox{$s=1,2,3,\,\cdots$} with never ending
   gauge invariances for 
gauge invariances is arising: We define  ${\bf R}^s={\bf d}$
 for \mbox{$s=1,2,3,\,\cdots$} 
so that  for each  gauge generator
 there exists an 
additional zero mode 
\beqa
{\bf R}^1\varepsilon^2_{o}=0, \quad &{\rm 
where}&\quad\varepsilon^2_{o}={\bf R}^2\varepsilon^2_{e} \no \\
{\bf R}^2\varepsilon^3_{e}=0, \quad &{\rm 
where}&\quad\varepsilon^3_{e}={\bf R}^3\varepsilon^3_{o}\no\\
\cdots\qquad&   &\qquad\cdots
\eeqa
due to the nilpotency ${\bf d}^2=0$.

\section{Gauge Fixing and  BV-Quantization}

In this section we straightforwardly apply the 
usual field theory 
BV-path integral quantization scheme \cite{BV2,Henn,Gomis}  
to the  Connes--Lott 2-point model:  
In addition to the original gauge field $\A$, which for notational
convenience  
we temporarily denote  by $\A\equiv\C^{-1}_{-1}$, we introduce  
  ghost fields 
$\C^{k}_s$, \, $\infty\ge s\ge -1,\,\,s\ge k\ge -1$ with $k \,\,{\rm
odd}$, as well 
as auxiliary ghost fields 
$\bar{\C}^{k}_s$, \, $\infty\ge s\ge 0,\,\,s\ge k\ge 0$ with $k \,\,
{\rm 
even}$ (see Fig. 1).

\begin{picture}(400,240)
\put(158,220){$\A\equiv{\cal C}_{-1}^{-1}$}
\put(171,218){\line(-3,-4){15}}
\put(150,190){\line(3,4){15}}
\put(200,190){\line(-3,4){20}}
\put(140,180){$\bar {\cal C}_{0}^{0}$}
\put(200,180){${\cal C}_{0}^{-1}$}
\put(123,154){\line(3,4){15}}
\put(141,178){\line(-3,-4){15}}
\put(183,154){\line(3,4){15}}
\put(201,178){\line(-3,-4){15}}
\put(230,150){\line(-3,4){20}}
\put(110,140){${\cal C}^{1}_1$}
\put(170,140){$\bar {\cal C}_{1}^{0}$}
\put(230,140){${\cal C}_{1}^{-1}$}
\put(93,114){\line(3,4){15}}
\put(111,138){\line(-3,-4){15}}
\put(153,114){\line(3,4){15}}
\put(171,138){\line(-3,-4){15}}
\put(213,114){\line(3,4){15}}
\put(231,138){\line(-3,-4){15}}
\put(260,110){\line(-3,4){20}}
\put(80,100){$\bar {\cal C}^2_{2}$}
\put(140,100){${\cal C}^{1}_2$}
\put(200,100){$\bar {\cal C}_{2}^{0}$}
\put(260,100){${\cal C}_{2}^{-1}$}
\put(63,74){\line(3,4){15}}
\put(81,98){\line(-3,-4){15}}
\put(123,74){\line(3,4){15}}
\put(141,98){\line(-3,-4){15}}
\put(183,74){\line(3,4){15}}
\put(201,98){\line(-3,-4){15}}
\put(243,74){\line(3,4){15}}
\put(261,98){\line(-3,-4){15}}
\put(290,70){\line(-3,4){20}}
\put(50,60){${\cal C}^{3}_3$}
\put(110,60){$\bar {\cal C}^2_{3}$}
\put(170,60){${\cal C}^{1}_3$}
\put(230,60){$\bar {\cal C}_{3}^{0}$}
\put(290,60){${\cal C}_{3}^{-1}$}
\put(170,35){.}
\put(170,30){.}
\put(170,25){.}
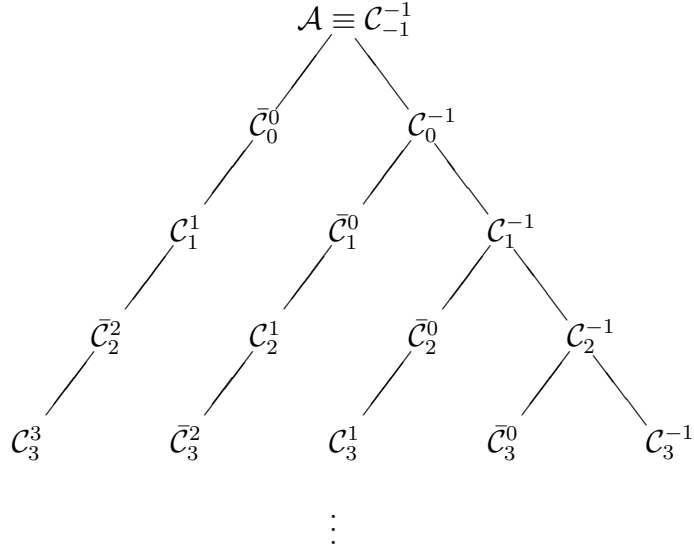
\put(70,5){\rm Figure 1.\ The Infinite Tower of Ghost Fields}
\end{picture}
\\
Furthermore we add Lagrange multiplier fields $\pi^{k}_s$, 
 $\infty\ge s\ge 1,\,s\ge k\ge 1$ with $k \,\,{\rm odd}$ and
$\bar{\pi}^{k}_s$,  $\infty\ge s\ge 0,\,s\ge k\ge 0$ with $k \,\, {\rm 
even}$. Finally we introduce antifields ${\C^{k}_s}^{*}$, ${\bar{\C}^{k}_s\,}^{*}$.
All the ghosts $\C^{k}_s$, $\bar{\C}^{k}_s$, multiplier fields
$\pi^{k}_s$, $\bar{\pi}^{k}_s$ and antifields 
${\C^{k}_s}^{*}$, ${\bar{\C}^{k}_s\,}^{*}$ are matrices which are 
even  for $s$ even and odd  for $s$ odd, respectively. We define all 
the  ghost fields $\C^{k}_s$, $\bar{\C}^{k}_s$ to be anti-Hermitean, all 
the multiplier fields
$\pi^{k}_s$, $\bar{\pi}^{k}_s$ to be Hermitean. When $s$ is taken to 
be odd the ghosts are 
bosonic whereas the multiplier fields are  fermionic; for $s$ even the 
ghosts are fermionic and  the multiplier fields are bosonic, respectively.

An important quantity for the construction of the BV-action is 
the commutator of (zero-stage) infinitesimal  gauge 
transformations $[\delta_{\varepsilon_{1}},\delta_{\varepsilon_{2}}]\A$, 
where   $\delta_{\varepsilon_{k}} 
\A=i {\bf R}^0\, \varepsilon_{k}$  with even matrices $\varepsilon_{k},
\,\,k=1,2$. It is 
easy to see that this commutator is vanishing. 
 The BV-action therefore obtains  as
\beq
S_{BV}=S_{inv}+S_{aux}-\langle{{\cal C}_{-1}^{-1}}^{*} \vert{\bf 
D}\,{\cal 
C}_{0}^{-1}\rangle -\sum_{s=1,3,5, \cdots}^{\infty} \langle{{\cal
C}_{s}^{-1}}^{*} 
\vert{\bf d}\,{\cal 
C}_{s+1}^{-1}\rangle
-i\sum_{s=0,2,4, \cdots}^{\infty} \langle{{\cal C}_{s}^{-1}}^{*} 
\vert{\bf d}\,{\cal 
C}_{s+1}^{-1}\rangle,
\eeq
where we denote by $S_{aux}$ the auxiliary field action 
\beq
S_{aux}=\sum_{k=0,2,4, \cdots}^{\infty} \,\sum_{s=k}^{\infty} 
\langle{\bar\pi}^{k}_{s}  \vert
{{{\bar{\cal C}}_{s}^{k}}\,}^{*}\rangle+
\sum_{k=1,3,5, \cdots}^{\infty} \,\sum_{s=k}^{\infty} 
\langle{{\C}^{k}_{s}}^{*} \vert
{{{{\pi}}_{s}^{k}}\,}\rangle.
\eeq
By $\mbox{\boldmath $\delta$}$ we denote a nilpotent
matrix coderivative operator\footnote
{\qquad $\mbox{\boldmath $\delta$} \, a =i\left( \ba{cc}
a_{12}-a_{21} & -a_{11}-a_{22}  \\
-a_{11}-a_{22} & -a_{12}+a_{21}
\ea \right)\qquad {\rm where } \qquad
a = \left( \ba{cc}
a_{11} & a_{12}  \\
a_{21} & a_{22} 
\ea \right), \quad a_{ij} \in \rm C. $} which 
is defined by \mbox{$\langle\mbox{\boldmath $\delta$} 
a_{o}\vert b_{e}\rangle= 
\langle a_{o}\vert{\bf d} b_{e}\rangle$} and  
\mbox{$\langle\mbox{\boldmath $\delta$} a_{e}\vert b_{o}\rangle= 
\langle a_{e}\vert{\bf d} b_{o}\rangle$}. It
allows to define   gauge fixing conditions 
\beqa
{{\mbox{\boldmath 
$\delta$}{\cal C}}_{s}^{k}}\,&=&0, \qquad
\infty\ge s\ge -1,\quad s\ge k\ge -1 \quad{\rm with} \,k \,\quad{\rm
odd}\no\\    
{\mbox{\boldmath 
$\delta$}\bar\C}^{k}_{s}&=&0, \qquad 
\infty\ge s\ge \,0,\quad\,\,\,\, s\ge k\ge \,0 \quad\,\,\,\,{\rm with} 
\,\, k \quad {\rm even},
\eeqa
which are similar to the  Feynman gauge in standard Yang--Mills theory. 
In the BV-approach we implement these gauge fixing conditions by
defining the gauge fixing fermion 
$\Psi=\Psi_{\delta}+\Psi_{\pi}$ by
\beqa
\Psi_{\delta}
&=&
\sum_{s=0,2,4, \cdots}^{\infty}\,\,\sum_{k=0,2,4, \cdots\, \, k\le s} \, 
\left(-\langle{\bar\C}^{k}_{s} \,\vert\,
{{{\mbox{\boldmath $\delta$}{\cal C}}_{s-1}^{k-1}}\,}\rangle +
\langle{{\mbox{\boldmath $\delta$}\bar\C}^{k}_{s+1} \,\vert\,
{{{\cal C}}_{s+2}^{k+1}}\,}\rangle 
\right.\no\\
&&\mbox{} 
\qquad\qquad\qquad\qquad\quad+
\left. 
i\langle{\bar\C}^{k}_{s+1} \,\vert\,
{{{\mbox{\boldmath $\delta$}{\cal C}}_{s}^{k-1}}\,}\rangle +
i\langle{{\mbox{\boldmath $\delta$}\bar\C}^{k}_{s} \,\vert\,
{{{\cal C}}_{s+1}^{k+1}}\,}\rangle\right),\no \\
\Psi_{\pi}&=&
\frac{1}{2} \sum_{s=0,2,4, \cdots}^{\infty}\,\,\sum_{k=0,2,4, \cdots\, 
\, k <  s} 
\left(\langle{\bar\C}^{k}_{s} \,\vert\,
{{{{\pi}}_{s}^{k+1}}\,}\rangle+
\,\langle{\bar\pi}^{k}_{s} \,\vert\,
{{{{\C}}_{s}^{k+1}}\,}\rangle
\right.\no\\
&&\mbox{} \qquad\qquad\qquad\qquad\quad+
\left. 
i\langle{\bar\C}^{k}_{s+1} \,\vert\,
{{{{\pi}}_{s+1}^{k+1}}\,}\rangle+
\,i\langle{\bar\pi}^{k}_{s+1} \,\vert\,
{{{{\C}}_{s+1}^{k+1}}\,}\rangle
\right)\no\\
&&\mbox{}+ \frac{1}{2}
\sum_{k=0,2,4, \cdots}^{\infty} \, 
\langle{\bar\C}^{k}_{k} \,\vert\,
{{{\bar{\pi}}_{k}^{k}}\,}\rangle.
\eeqa
We eliminate the
antifields  by using the gauge fixing fermion 
$\Psi$ via
\beq
\langle{\C^{k}_s}^{*}\vert=\frac{\partial\Psi}{\partial \vert 
\C^{k}_s\rangle}, \quad
\vert{\bar\C^{k}_s\,}^{*}\rangle=\frac{\partial \Psi}{\partial 
\langle\bar\C^{k}_s\vert}, 
\eeq
so that the gauge fixed action $S_{\Psi}$ reads
\beqa
S_{\Psi}&=&S_{inv}- i \langle {\bar \C}_{0}^{0} \,\vert\,
\mbox{\boldmath $\delta$} 
{\bf D}\,
{\C}_{0}^{-1}\rangle -i
\sum_{s=1,3,5, \cdots}^{\infty} \langle{\bar \C}_{s+1}^{0}
 \,\vert\, \mbox{\boldmath $\delta$}{\bf d}\,{\C}_{s+1}^{-1}\rangle +
 \sum_{s=0,2,4, \cdots}^{\infty} \langle{\bar \C}_{s+1}^{0}
 \,\vert\, \mbox{\boldmath $\delta$}{\bf d}\,{\C}_{s+1}^{-1}\rangle 
 \no\\
&&\mbox{}+ 
\sum_{k=0,2,4, \cdots}^{\infty} \,\, \sum_{s=k+1,\, odd}^{\infty}
 \left(i \langle{{{{\bar\pi}}_{s}^{k}}} \,\vert\, 
{\pi}^{k+1}_{s}\rangle 
+\langle{\bar\pi}^{k}_{s}  \, 
\vert \,(i  \mbox{\boldmath $\delta$}{{{{\C}}_{s-1}^{k-1}}\,}+
 \,{\bf d}{\C}^{k+1}_{s+1})\rangle\no \right.\\
&&\mbox{}\left. \qquad\qquad\qquad\qquad\quad+\langle(i
\mbox{\boldmath $\delta$}{{{{\bar\C}}_{s-1}^{k}}\,}\,-
{\bf d}{\bar\C}^{k+2}_{s+1})\vert\,  
{{\pi}}_{s}^{k+1}\rangle\right)\no\\
&&\mbox{}+ 
\sum_{k=0,2,4, \cdots}^{\infty}\,\,  \sum_{s=k+2,\, even}^{\infty}
 \left( \langle{{{{\bar\pi}}_{s}^{k}}} \,\vert\, 
{\pi}^{k+1}_{s}\rangle 
+\langle{\bar\pi}^{k}_{s}  \, 
\vert \,(-  \mbox{\boldmath $\delta$}{{{{\C}}_{s-1}^{k-1}}\,}+
 \,i{\bf d}{\C}^{k+1}_{s+1})\rangle \right. \no\\
&&\mbox{}\left. \qquad\qquad\qquad\qquad\quad+\langle(
\mbox{\boldmath $\delta$}{{{{\bar\C}}_{s-1}^{k}}\,}\,+
i{\bf d}{\bar\C}^{k+2}_{s+1})\vert\,  
{{\pi}}_{s}^{k+1}\rangle\right)\no\\
&&\mbox{}+\sum_{k=0,2,4, \cdots}^{\infty} \,
\langle{\bar\pi}^{k}_{k}  \,\vert\, 
( - \mbox{\boldmath $\delta$}{{{{\C}}_{k-1}^{k-1}}\,}+
i{\bf d}{\C}^{k+1}_{k+1} \,+\frac{1}{2}
{{{{\bar\pi}}_{k}^{k}}})\rangle. 
\eeqa
We can now  eliminate the Lagrange multiplier fields $\pi^{k}_s$ 
and $\bar{\pi}^{k}_s$  and  arrive at
\beqa
S_{\Psi}\longrightarrow &&S_{inv}+ \frac{1}{2}\langle\A \,\vert \,
{\bf d}\mbox{\boldmath $\delta$} 
\,
\A\rangle-i
\langle{\bar \C}_{0}^{0}\,\vert\,(
\mbox{\boldmath $\delta$} 
{\bf D}\,+{\bf d}\mbox{\boldmath $\delta$})\,
{\C}_{0}^{-1}\rangle \no\\
&&\mbox{}-i\sum_{s=1,3,5, \cdots}^{\infty} \langle{\bar \C}_{s+1}^{0}
 \,\vert\, (\mbox{\boldmath $\delta$}{\bf d}+{\bf d}\mbox{\boldmath 
 $\delta$})\,{\C}_{s+1}^{-1}\rangle\no\\
&&\mbox{} +\sum_{s=0,2,4, \cdots}^{\infty} \langle{\bar \C}_{s+1}^{0}
 \,\vert\, (\mbox{\boldmath $\delta$}{\bf d}+{\bf d}\mbox{\boldmath 
 $\delta$})\,{\C}_{s+1}^{-1}\rangle \no\\
&&\mbox{}-
i\sum_{k=0,2,4, \cdots}^{\infty}\, \,\sum_{s=k+1, \, odd}^{\infty}
\langle{\bar \C}_{s+1}^{k+2}\,\vert\,(
\mbox{\boldmath $\delta$} 
{\bf d}\,+{\bf d}\mbox{\boldmath $\delta$})\,
{\C}_{s+1}^{k+1}\rangle\no\\
&&\mbox{}+
\sum_{k=0,2,4, \cdots}^{\infty}\, \,\sum_{s=k+2, \, even}^{\infty}
\langle{\bar \C}_{s+1}^{k+2}\,\vert\,(
\mbox{\boldmath $\delta$} 
{\bf d}\,+{\bf d}\mbox{\boldmath $\delta$})\,
{\C}_{s+1}^{k+1}\rangle\no\\
&&\mbox{} +\frac{1}{2}\sum_{k=0,2,4, \cdots}^{\infty}  
\langle
{\C}_{k+1}^{k+1}\,\vert\,(
\mbox{\boldmath $\delta$} 
{\bf d}\,+{\bf d}\mbox{\boldmath $\delta$})\,
{\C}_{k+1}^{k+1}\rangle 
.
\eeqa
All the higher-stage ghost  contributions  can be integrated away 
without any effect as
$\mbox{\boldmath $\delta$} 
{\bf d}\,+{\bf d}\mbox{\boldmath $\delta$}=4 \cdot{\bf 1}$ and we 
simply
obtain
\beq
\label{fixfinal}
S_{\Psi} \longrightarrow S_{inv}+ \frac{1}{2}\langle\A \,\vert \,
{\bf d}\mbox{\boldmath $\delta$} 
\,
\A\rangle-
i\langle{\bar \C}_{0}^{0}\,\vert\,(
\mbox{\boldmath $\delta$} 
{\bf D}\,+{\bf d}\mbox{\boldmath $\delta$})\,
{\C}_{0}^{-1}\rangle
\eeq
We see
that  
the  gauge fixed action contains the  invertible quadratic part 
$2\langle\A \,\vert \,
\A\rangle$ for the gauge field, as well as 
$-4i \langle{\bar \C}_{0}^{0}\,\vert\,
{\C}_{0}^{-1}\rangle$
for  the  ${\bar \C}_{0}^{0}$, 
${\C}_{0}^{-1}$ ghost fields. 

\section{The Gribov Problem}

The Yang--Mills theory on the 2-point space suffers from a Gribov 
problem \cite{Gribov} even for the abelian $U(1)\times U(1)$ case. 
This can be demonstrated easily by recasting the ghost  part of the 
gauge fixed action (\ref{fixfinal}) into the form 
\beq
\langle{\bar \C}_{0}^{0}\,\vert \,
(\mbox{\boldmath $\delta$} 
{\bf D}\,+{\bf d}\mbox{\boldmath $\delta$})\,
{\C}_{0}^{-1}\rangle =
\ba{c} ({\bar c}_{1}\quad {\bar c}_{2}){}\ea 
\left( \ba{cc}
4+\phi+\bar \phi & -\phi-\bar \phi  \\
-\phi-\bar \phi & 4+\phi+\bar \phi 
\ea \right)
\left( \ba{c}
{c}_{1}  \\
{c}_{2} 
\ea \right),
\eeq
where we introduced the component ghosts fields 
${\bar c}_{1},\,{\bar c}_{2}$ and $c_{1},\,c_{2}$ which 
are the diagonal elements of
 ${\bar \C}_{0}^{0}$ and 
${\C}_{0}^{-1}$, respectively. The Faddeev--Popov matrix 
$M_{FP}$
\beq
M_{FP}=\left( \ba{cc}
4+\phi+\bar \phi & -\phi-\bar \phi  \\
-\phi-\bar \phi & 4+\phi+\bar \phi 
\ea \right)
\eeq
has a vanishing determinant
 for $2+\phi+\bar\phi=0$ which forces
 $\phi=u+i v$ to lie on the line  $u=-1$. 
We note the  distinguished value  $\phi=-1$, which we discussed 
previously by demanding
  the action 
 to be 
 maximal, see (\ref{max}). Now  this value arises by 
 inserting the gauge fixing 
 condition $\mbox{\boldmath $\delta$} 
\,\A=0$ into the Faddeev--Popov determinant $det \,M_{FP}$.
 
 We  observe  that the   classical 
 action $S_{inv}$   not only has an invariance under the (rigid) gauge 
 transformations (\ref{old}), but also under the discrete charge
conjugation 
 operation (conveniently expressed in terms of the Higgs fields 
 $H,\,\bar H$)
 \beq
 H \longrightarrow -\bar H,\qquad \bar H \longrightarrow -H.
 \eeq
 After the spontaneous symmetry breakdown this discrete symmetry 
guarantees that  the minima of the action are degenerated. 
 In the quantum case, however, due to the Gribov problem,  these
discrete jumps 
 no longer are allowed and the quantum corrections to the action will
lift 
 the classical degeneracy of the minima.

\section{Connes'  Real Spectral Triple Formulation}

The formulation of the Yang--Mills theory model on the 2-point space 
in terms of Connes's real spectral triple approach  
proceeds by specifying the spectral triple $({\mathbf A}, {\cal H}, D)$ 
together with the antilinear isometry ${\cal J}$,  fulfilling   a set of 
specific properties \cite{gravity,reality}. We represent the elements 
$a=(a_{1},a_{2},a_{3})$ of  the algebra ${\mathbf 
A}=C \oplus C \oplus C$, as well as $D$ and $\cal J$,  by specific 
$4\times 4$ 
 matrices; the Hilbert space ${\cal H}$ simply is 
 $C^4$. Specifically we have
\beq
a= \left( \ba{cccc}
 a_{1} & 0 & 0 & 0 \\
0 &  a_{2} & 0 & 0\\
0 & 0 & a_{3} & 0 \\
0 & 0 & 0 & a_{3}
\ea \right)
\quad
D=\left( \ba{cccc}
 0 & 1 & 0 & 0 \\
1 & 0 & 0 & 0\\
0 & 0 & 0 & 1 \\
0 & 0 & 1 & 0
\ea \right)
\quad
{\cal J}=\left( \ba{cccc}
 0 & 0 & 1 & 0 \\
0 & 0 & 0 &1\\
1 & 0 & 0 & 0 \\
0 &1 & 0 & 0
\ea \right)\circ\, c.c.
\eeq
where c.c. denotes complex conjugation. As an example one sees that for
$a,\,b \in {\mathbf 
A}$ the differential  
1-form \,$\omega_{1}=a \, [i \,D,b]$  is given by
\beq
\omega_{1}=i\left( \ba{cccc}
 0 & a_{1}(b_{2}-b_{1}) & 0 & 0 \\
a_{2}(b_{1}-b_{2}) & 0 & 0 &0\\
0 & 0 & 0 & 0 \\
0 &0 & 0 & 0
\ea \right).
\eeq
We recognize that apart of  irrelevant zeros  in the upper 
right, lower left and lower right matrix corners of the differential 
forms our previous discussion of the 
 gauge symmetries, the gauge fixing and the ghost structure proves 
 right as well. 
 
 We conclude that the quantization of the Yang--Mills theory model on
the 
 2-point space within the Connes--Lott 
 scheme and within
  Connes' real 
 spectral triple  approach are equivalent; the model reveals infinite
reducibility 
and is afflicted with the  Gribov problem.
\\{\bf \it{Note added:}}  After finishing our paper a related article 
\cite{Hauss}
appeared. One of its main purposes is to analyse in depth the counting
argument 
of Feynamn diagrams in the presence of spontaneous symmetry breaking. 
Concerning the gauge
fixing procedure and the introduction of ghost fields, however, the
investigation 
 appears to be incomplete.

\section*{Acknowledgement}

I thank John Madore for suggesting to study the
Yang--Mills theory model on the 2-point space. I am very grateful to 
Harald Grosse and Raimar Wulkenhaar for innumerable helpful discussions.

\end{document}